\documentstyle[sprocl]{article}

\def\Journal#1#2#3#4{{#1} {\bf #2}, #3 (#4)}

\def\NPB{{\em Nucl. Phys.} B}
\def\PLB{{\em Phys. Lett.}  B}

\def\PRD{{\em Phys. Rev.} D}

\def\be{\begin{equation}}
\def\ee{\end{equation}}
\def\bea{\begin{eqnarray}}
\def\eea{\end{eqnarray}}
\def\DG{\Delta\Gamma_{B_s}}
\newcommand{\beq}{\begin{equation}}
\newcommand{\eeq}{\end{equation}}
\newcommand{\beqa}{\begin{eqnarray}}
\newcommand{\eeqa}{\end{eqnarray}}

\begin{document}
 \begin{flushright}
  \begin{tabular}{l}
  SLAC-PUB-7258\\
  hep-ph/9609214\\
  August 1996
  \end{tabular}
  \end{flushright}
  \vskip0.5cm

\title{LIFETIME DIFFERENCE OF $B_s$ MESONS AND ITS 
IMPLICATIONS~\footnote{
Research supported 
by the Department of Energy under contract DE-AC03-76SF00515.}
}

\author{ M. BENEKE }

\address{Stanford Linear Accelerator Center,\\ 
Stanford University, Stanford, CA 94309, U.S.A.}

\vspace{1.5cm}

\maketitle\abstracts{
We discuss the calculation of 
the width difference $\Delta\Gamma_{B_s}$ between the $B_s$ mass 
eigenstates to next-to-leading order in the heavy quark expansion. 
$1/m_b$-corrections are estimated to reduce the leading order result 
by typically $30\%$. The error of the 
present estimate $(\Delta\Gamma/\Gamma)_{B_s}=0.16^{+0.11}_{-0.09}$ 
could be substantially improved by pinning down the value of 
$\langle \bar{B}_s |(\bar b_is_i)_{S-P}(\bar b_js_j)_{S-P}| 
B_s\rangle$ and an accuracy of $10\%$ in
$(\Delta\Gamma/\Gamma)_{B_s}$ should eventually be reached. 
We briefly mention strategies to measure $(\Delta\Gamma/\Gamma)_{B_s}$, 
and its implications for
constraints on $\Delta M_{B_s}$, CKM parameters and the 
observation of CP violation in untagged $B_s$ samples.}

\vspace*{2cm}

\begin{center}
{\em To appear in the Proceedings of the \\
     28th International Conference\\
     on High Energy Physics,\\
     Warsaw, Poland, 25-31 July 1996}
\end{center}

\newpage

\section{Introduction}

Mixing phenomena in neutral $B$ meson systems provide us with an
important probe of standard model flavordynamics and its 
interplay with the strong interaction. As is well-known, non-zero 
off-diagonal elements of the mixing matrix in the flavor basis  
$\{|B_s\rangle,\, |\bar B_s\rangle\}$ are generated in second order 
in the weak interaction through `box diagrams'. In the 
$B_s$ system\footnote{For $B_d$ mesons there is further CKM suppression 
and their lifetime difference will not be considered here.}
the off-diagonal elements obey the pattern
\begin{equation}
\left|\frac{\Gamma_{12}}{M_{12}}\right|\sim {\cal O}\!\left(\frac{m_b^2}
{m_t^2}\right).
\end{equation}
The mass and lifetime 
difference between eigenstates are given by (`H' for `heavy', 
`L' for `light')
\begin{equation}\label{delmex}
\Delta M_{B_s} \equiv M_H-M_L=2 \,|M_{12}|,\\[0.1cm]
\end{equation}
\begin{equation}\label{delgex}
\Delta \Gamma_{B_s} \equiv \Gamma_L-\Gamma_H = -\frac{2\,\mbox{Re}\,(
M_{12}^*\Gamma_{12})}{|M_{12}|}\approx -2 \Gamma_{12},
\end{equation}
up to very small corrections (assuming standard model CP violation).  
Anticipating the magnitudes of the eigenvalues, 
we have defined both $\Delta M_{B_s}$ and $\Delta\Gamma_{B_s}$ 
to be positive. Note that the lighter state is CP even and decays 
more rapidly than the heavier state. 

The lifetime difference is an interesting quantity in several 
respects. Contrary to the neutral kaon system, it is calculable 
by short-distance methods and directly probes the spectator quark 
dynamics which generates lifetime differences among all $b$ hadrons. 
If the mass difference $\Delta M_{B_s}$ turns out to be large, 
the lifetime difference also tends to be large and may well be the 
first direct observation of mixing for $B_s$ mesons. If 
$\Delta\Gamma_{B_s}$ is sizeable, CP violation in the 
$B_s$ system can be observed without 
flavor-tagging \cite{DUN1}.

The following sections summarize the calculation of Ref.~\cite{BBD} 
and discuss some of the implications of a non-zero $\Delta\Gamma_{B_s}$. 

\section{Heavy quark expansion of $\DG$}

The mass difference is dominated by the top-quark box diagram, which 
reduces to a local $\Delta B=2$ vertex on a momentum scale smaller than 
$M_W$. The lifetime difference, on the other hand, is generated by 
real intermediate states and is not yet local on this scale. But  
the $b$ quark mass $m_b$ provides an additional short-distance 
scale that leads to a large energy release 
(compared to $\Lambda_{QCD}$) into the intermediate states. Thus, 
at typical hadronic scales the decay is again a local process. The 
lifetime difference can then be treated by the same operator product 
expansion that applies to the average $B_s$ lifetime and 
other $b$ hadrons \cite{BIG}. 

Summing over all intermediate states, the off-diagonal element 
$\Gamma_{21}$ of the decay width matrix is given by
\begin{equation}\label{g21t}
\Gamma_{21}=\frac{1}{2M_{B_s}}
\langle\bar B_s|\,\mbox{Im}\,i\!\int\!\!d^4xT\,{\cal H}_{eff}(x)
{\cal H}_{eff}(0)|B_s\rangle 
\end{equation}
with 
\beqa\label{heff}
{\cal H}_{eff}&=&\frac{G_F}{\sqrt{2}}V^*_{cb}V_{cs}
\big(C_1(\mu) (\bar b_ic_j)_{V-A}(\bar c_js_i)_{V-A}  
\nonumber\\
&& +\,C_2(\mu) (\bar b_ic_i)_{V-A}(\bar c_js_j)_{V-A}\big).
\eeqa
Cabibbo suppressed and penguin operators in $
{\cal H}_{eff}$ have not been written 
explicitly. In leading logarithmic approximation, the Wilson coefficients 
are given by 
$C_{2,1}=(C_+\pm C_-)/2$, where
\beq
C_+(\mu)\!=\!\left[\frac{\alpha_s(M_W)}{\alpha_s(\mu)}\right]^{6/23}\!\!\!\!
C_-(\mu)\!=\!\left[\frac{\alpha_s(M_W)}{\alpha_s(\mu)}\right]^{-12/23}
\end{equation}
\noindent and $\mu$ is of order $m_b$.

 The heavy quark expansion expresses $\DG$ as a series in local 
$\Delta B=2$-operators. In the following we keep $1/m_b$-corrections to 
the leading term in the expansion. Keeping these terms fixes various 
ambiguities of the leading order calculation, such as whether the 
quark mass $m_b$ or meson mass $M_{B_s}$ should be used, and establishes 
the reliability of the leading order expression obtained in 
Ref.~\cite{LO,VOL}. Compared to the `exclusive approach' 
pursued in Ref.~\cite{ALE93} that 
adds the contributions to $\DG$ from individual intermediate states, the 
inclusive approach is model-independent. The operator product expansion 
provides a systematic approximation in $\Lambda_{QCD}/m_b$, but it 
relies on the assumption of `local duality'. The accuracy to which 
one should expect duality to hold is difficult to quantify, except for 
models \cite{CHI96} and eventually by comparison with data. We shall 
assume that duality violations will be less than 10\% for $\DG$. 

To leading order in the heavy quark expansion, the long distance 
contributions to $\DG$ are parameterized by the matrix elements of 
two dimension six operators
\begin{eqnarray}\label{qqs}
Q &=& (\bar b_is_i)_{V-A}(\bar b_js_j)_{V-A},\\[0.1cm]
Q_S\!\!&=& (\bar b_is_i)_{S-P}(\bar b_js_j)_{S-P} 
\end{eqnarray}
between a $\bar{B}_s$ and $B_s$ state. We write these matrix elements 
as 
\begin{eqnarray}\label{meqs}
\langle Q\rangle &=& \frac{8}{3}\, f^2_{B_s}M^2_{B_s}\,B,
\\[0.1cm]
\langle Q_S\rangle\!\! &=& - \frac{5}{6} \,f^2_{B_s}M^2_{B_s}
\frac{M^2_{B_s}}{(m_b+m_s)^2}\,B_S,
\end{eqnarray}
where $f_{B_s}$ is the $B_s$ decay constant. 
The `bag' parameters $B$ and $B_S$ are defined such that 
$B=B_S=1$ corresponds to factorization. $B$ also appears in the 
mass difference, while $B_S$ is specific to $\DG$. 

The matrix elements of these operators are not independent of $m_b$. 
Their $m_b$-dependence could be extracted with the help of 
heavy quark effective theory. There seems to be no gain 
in doing so, since the number of independent 
nonperturbative parameters is not reduced even at leading order 
in $1/m_b$ and since we work to subleading order in $1/m_b$ even 
more parameters would appear. The 
matrix elements of the local $\Delta B=2$-operators should therefore be 
computed in `full' QCD, for instance on the lattice. 

Including $1/m_b$-corrections, the width difference is found to be
\begin{eqnarray}\label{tres}
\DG &=& \frac{G^2_F m^2_b}{12\pi M_{B_s}}(V^*_{cb}V_{cs})^2 
\sqrt{1-4z}
\nonumber\\
&&\hspace*{-1.7cm}\,\cdot\bigg[\left((1-z)K_1+
\frac{1}{2}(1-4z)K_2\right)\langle Q 
\rangle\\ 
&&\hspace*{-1.7cm}\,+\,(1+2z)\left(K_1-K_2\right)\langle Q_S\rangle  + 
\hat{\delta}_{1/m} + \hat{\delta}_{rem} \bigg],
\nonumber
\end{eqnarray}
where $z=m^2_c/m^2_b$ and
\begin{equation}\label{k1k2}
K_1=N_c C^2_1+2C_1 C_2\qquad K_2=C^2_2 .
\end{equation}
The $1/m_b$-corrections are summarized in 
\begin{eqnarray}\label{oneoverm}
\hat{\delta}_{1/m} &=& (1+2 z)\Big[K_1\,(-2\langle R_1\rangle 
-2\langle R_2\rangle) 
+\,K_2\,(\langle R_0\rangle -2\langle \tilde{R}_1
\rangle -2\langle \tilde{R}_2\rangle)\Big]
\nonumber\\
&&-\,\frac{12 z^2}{1-4 z}\Big[K_1\,(\langle R_2\rangle 
+2\langle R_3\rangle) 
+\,K_2\,(\langle \tilde{R}_2\rangle 
+2\langle \tilde{R}_3\rangle)\Big] .
\end{eqnarray}
The operators $R_i$ and $\tilde{R_i}$ involve derivatives on quark 
fields or are proportional to the strange quark mass $m_s$, which we 
count as $\Lambda_{QCD}$. For instance,
\begin{eqnarray}
\label{r0qt}
R_1\!&=&\!\frac{m_s}{m_b}(\bar b_is_i)_{S-P}(\bar b_js_j)_{S+P},\\
\label{rrt2}
R_2\!&=&\!\frac{1}{m^2_b}(\bar b_i {\overleftarrow D}_{\!\rho}
D^\rho s_i)_{V-A}( \bar b_j s_j)_{V-A}.
\label{rrt3}
\end{eqnarray}
The complete set can be found in Ref.~\cite{BBD}. Operators with gluon 
fields contribute only at order $(\Lambda_{QCD}/m_b)^2$. Since the 
matrix elements of the $R_i$, $\tilde{R}_i$ are $1/m_b$-suppressed 
compared to those of $Q$ and $Q_S$, we estimate them in the factorization 
approximation, assuming factorization at a scale of order $m_b$ (A smaller 
scale would be preferable, but would require us to calculate the 
anomalous dimension matrix.). Then all matrix elements can be 
expressed in terms of quark masses and the $B_s$ mass and decay constant. 
No new nonperturbative parameters enter at order $1/m_b$ in this 
approximation.

The term $\hat{\delta}_{rem}$ denotes the contributions from 
Cabibbo-suppressed decay modes and pengiun operators. They can be 
estimated \cite{BBD} to be below $\pm 3\%$ and about $-5\%$, 
respectively, relative to the leading order contribution. We neglect 
this term in the following numerical analysis. 

\section{Numerical estimate}

\begin{table}[t]
\addtolength{\arraycolsep}{-0.01cm}
\renewcommand{\arraystretch}{1.3}
\caption{\label{table1}
Dependence of $a$, $b$ and $c$ on the $b$-quark mass (in GeV) 
and renormalization 
scale for fixed values of all other short-distance parameters. The last 
column gives $(\Delta\Gamma/\Gamma)_{B_s}$ for $B=B_S=1$ (at 
given $\mu$), $f_{B_s}=210\,$MeV.}
$$
\begin{array}{|c|c||c|c|c|c|}
\hline
m_b & \mu & a & b & c & 
(\Delta\Gamma/\Gamma)_{B_s} \\ 
\hline\hline
4.8 & m_b & 0.009 & 0.211 & -0.065 & 0.155 \\ \hline
4.6 & m_b & 0.015 & 0.239 & -0.096 & 0.158 \\ \hline
5.0 & m_b & 0.004 & 0.187 & -0.039 & 0.151 \\ \hline
4.8 & 2 m_b & 0.017 & 0.181 & -0.058 & 0.140 \\ \hline
4.8 & m_b/2 & 0.006 & 0.251 & -0.076 & 0.181 \\ \hline
\end{array}
$$
\end{table}

It is useful to separate the dependence on the long-distance parameters 
$f_{B_s}$, $B$ and $B_S$ and write $(\Delta\Gamma/\Gamma)_{B_s}$ as 
\begin{equation}
\left(\frac{\Delta\Gamma}{\Gamma}\right)_{B_s} = 
\Big[a B + b B_S + c\Big]\left(\frac{f_{B_s}}{210\,\mbox{MeV}}\right)^2,
\end{equation}
where $c$ incorporates the explicit $1/m_b$-corrections. In the 
numerical analysis, we express $\Gamma_{B_s}$ as the theoretical value of 
the semileptonic width divided by the semileptonic branching ratio. 
The following parameters are kept fixed: $m_b-m_c=3.4\,$GeV, 
$m_s=200\,$MeV, $\Lambda^{(5)}_{LO}=200\,
$MeV, $M_{B_s}=5.37\,$GeV, $B(B_s\to X e\nu)=10.4\%$. Then $a$, $b$ and 
$c$ depend only 
on $m_b$ and the renormalization scale $\mu$. For some values of 
$m_b$ and $\mu$, the coefficients $a$, $b$, $c$ are listed in 
Tab.~\ref{table1}. For a central choice of parameters, which we take 
as $m_b=4.8\,$GeV, $\mu=m_b$, $B=B_S=1$ and $f_{B_s}=210\,$MeV, we 
obtain $(\Delta\Gamma/\Gamma)_{B_s} = 0.220 - 0.065 = 0.155$, where the 
leading term and the $1/m_b$-correction are
separately quoted. We note that the $V-A$ `bag' parameter $B$ has a very 
small coefficient and is practically negligible. The $1/m_b$-corrections 
are not small and decrease the prediction for $\DG$ by about 30\%.

The largest theoretical uncertainties arise from the decay constant 
$f_{B_s}$ and the second `bag' parameter $B_S$. In the large-$N_c$ 
limit, one has $B_S=6/5$, while estimating $B_S$ by keeping the 
logarithmic dependence on $m_b$ (but not $1/m_b$-corrections as required 
here for consistency) and assuming factorization at the scale $1\,$GeV 
gives \cite{VOL} $B_S=0.88$. $B_S$ has never been studied by either 
QCD sum rules or lattice methods. In order to estimate the range of 
allowed $\DG$ conservatively, we vary $B_S=1\pm0.3$, 
$f_{B_s}=(210\pm30)\,$MeV and obtain
\begin{equation}\label{dgnum1}
\left(\frac{\Delta\Gamma}{\Gamma}\right)_{B_s} = 0.16^{+0.11}_{-0.09}.
\end{equation}
This estimate could be drastically improved with improved knowledge of 
$B_S$ and $f_{B_s}$.

\section{Measuring $\DG$}

In principle, both $\Gamma_L$ and $\Gamma_H$ can be measured by 
following the time-dependence of flavor-specific modes \cite{DUN1}, 
such as $\bar{B}_s\to D_s l\nu$, given by 
\beq
e^{-\Gamma_H t}+e^{-\Gamma_L t} .
\eeq
In practice, this is a tough measurement. Alternatively, since the 
average $B_s$ lifetime is predicted~\cite{BBD} to be equal to the 
$B_d$ lifetime within $1\%$, it is sufficient to measure either 
$\Gamma_L$ or $\Gamma_H$. 

 The two-body decay $B_s\to D_s^+ D_s^-$ has a pure CP even final state 
and measures $\Gamma_L$. Since $D^0$ and $D^{\pm}$ do not decay into 
$\phi$ as often as $D_s$, the $\phi\phi X$ final state tags a 
$B_s$-enriched $B$ meson sample, whose decay distribution informs us  
about $\Gamma_L$.

 A cleaner channel is $B_s\to J/\psi\phi$, which has both CP even and 
CP odd contributions. These could be disentangled by studying the 
angular correlations \cite{DIG96}. In practice, this might not be 
necessary, as the CP even contribution is expected \cite{ALE93} 
to be dominant by 
more than an order of magnitude. In any case, the 
inequality
\beq \Gamma_L \geq 1/\tau(B_s\to J/\psi\phi)
\eeq
holds. CDF \cite{MES} has fully reconstructed 58 $B_s\to J/\psi\phi$ 
decays from 
run Ia+Ib and determined $\tau(B_s\to J/\psi\phi)= 
1.34^{+0.23}_{-0.19}\pm 0.05\,$ps. Together with 
$\tau(B_d)=1.54\pm 0.04\,$ps, assuming equal average $B_d$ and $B_s$ 
lifetimes, this yields 
\begin{equation}
\left(\frac{\Delta\Gamma}{\Gamma}\right)_{B_s}\geq 0.3\pm 0.4,
\end{equation} 
which still fails to be significant.
 
 In the Tevatron run II, as well as at HERA-B, one expects $10^3-10^4$ 
reconstructed $J/\psi\phi$, which will give a precise measurement of 
$\DG$.

\section{Implications of non-zero $\DG$}

\subsection{CKM elements}

Once $\DG$ is measured (possibly before $\Delta M_{B_s}$ is measured!), 
an alternative route to obtain the mass difference could use this 
measurement combined with the theoretical prediction for $(\Delta M/
\Delta\Gamma)_{B_s}$ \cite{DUN1,BP}. The decay constant $f_{B_s}$ drops 
out in this ratio, as well as the dependence on CKM elements, since 
$|(V_{cb}V_{cs})/(V_{ts}V_{tb})|^2= 1\pm 0.03$ by CKM unitarity. However, 
the dependence on long-distance matrix elements does not 
cancel even at leading order in $1/m_b$ and the prediction depends on 
the ratio of `bag' parameters $B_S/B$, which is not very well-known 
presently. We obtain $\Delta\Gamma/\Delta M=(5.6\pm 2.6)\cdot 10^{-3}$, 
where the largest error ($\pm 2.3$) arises from varying $B_S/B$ between 
0.7 and 1.3.

When lattice measurements yield an accurate value of $B_S/B$ as well 
as control over the $SU(3)$ flavor-symmetry breaking in $B f_B^2$, the 
above indirect determination of $\Delta M_{B_s}$ in conjunction with 
the measured mass difference in the $B_d$ system provides an
alternative way of determining the CKM ratio $|V_{ts}/V_{td}|$,
especially if the latter is around its largest currently
allowed value. In contrast, the ratio 
$\Gamma(B\to K^*\gamma)/\Gamma(B\to\{\varrho,\omega\}\gamma)$
is best suited for extracting small $|V_{ts}/V_{td}|$ ratios,
provided the long distance effects can be sufficiently well
understood.

\subsection{CP violation}

The existence of a non-zero $\DG$ allows the observation of 
mixing-induced CP asymmetries without tagging the initial $B_s$ or 
$\bar{B}_s$ \cite{DUN1,DF}. These measurements are difficult, but the 
gain in statistics, when tagging is obviated, makes them worthwhile  
to be considered. The mass difference drops out in the time 
dependence of untagged 
samples, which is given by 
\beq
A_+ (e^{-\Gamma_L t}+e^{-\Gamma_H t}) 
+ A_-(e^{-\Gamma_L t}-e^{-\Gamma_H t}).
\eeq
$A_-$ carries CKM phase information even in the absence of direct 
CP violation.
 
In combination with an analysis of angular distributions, a measurement 
of the CKM angle $\gamma$ from exclusive $B_s$ decays governed by the 
$\bar{b}\to c\bar{c}\bar{s}$ or $\bar{b}\to\bar{c}u\bar{s}$ transition 
can be considered \cite{DF}.

\section*{Acknowledgements}

I am grateful to my collaborators G.~Buchalla and I.~Dunietz for 
sharing their insights into problems related to this work with me.

\end{document}